# Simulations of charge-breeding processes in ECRIS

V. Mironov, S. Bogomolov, A. Bondarchenko, A. Efremov, and V. Loginov

*Joint Institute for Nuclear Research, Flerov Laboratory of Nuclear Reactions, Dubna, Moscow Reg. 141980, Russia*

**Abstract**. *Charge-breeding processes in Electron Cyclotron Resonance Ion Sources are numerically simulated by using the target helium plasma parameters obtained with NAM-ECRIS code. Breeding efficiency is obtained as a function of 1+ ion injection energy for some alkali ion beams. Time dependencies of extracted ions are calculated; typical times for reaching saturation in currents are in the range of few tens of milliseconds. Role of charge-exchange processes in breeding of ions is discussed. Recycling of ions on the source walls is shown to be important.*

## 1. Introduction

The Isotope-Separation Online (ISOL) method is a technique to produce Radioactive Ion Beams (RIB) for fundamental and applied researches. The method requires a driver accelerator to produce radioactive atoms essentially at rest and a post-accelerator to accelerate RIB to energies of interest. The accelerators are coupled by a system that consists of a production target, a singly charged ion (1+) source and a charge breeder that produces RIB in the charge state needed for the post-accelerator in (1+)→(n+) process. Electron Cyclotron Resonance Ion Sources (ECRIS) are often used as the charge-breeders in the modern ISOL installations [1]. The key parameters of the charge-breeders are efficiency and purity of producing the beams in the requested charge state; breeder optimization requires careful preparation both of the singly charged ion beam and of the target plasma where ions are captured and ionized into higher charge states. Numerical simulations of breeding process can be useful for increasing the breeders' efficiency.

Capture of the externally injected singly charged ions by ECRIS plasma is numerically studied in a series of publications of A. Galatà et al. [2,3,4]. There, experimentally observed changes of the transmission of the (1+) ion beam through the plasma with varying the beam initial energy are accurately reproduced; some important aspects, such as efficiency and characteristic times of the breeding process, were not studied.

We apply *NAM-ECRIS* code [5, 6, 7] to investigate the dynamics of the (1+)→(n+) process. At this stage our purpose is to reveal basic features, and we use parameters of DECRIS-SC2 18-GHz source [8], not of any ECRIS specially designed as breeder. From point of view of ionization efficiency, conventional ECRIS and charge-breeders are essentially the same devices [9]. The major difference is that breeders are equipped with a grounded tube to control the (1+) beam injection, which results in different geometry of the injection side.

## 2. Target plasma parameters

The *NAM-ECRIS* code uses the Particle-in-Cell Monte-Carlo Collisions approach for simulations of the ion dynamics in ECRIS plasma. In the present configuration, the model describes ECRIS plasma as dense plasmoid limited by (relativistically broadened) ECR surface, at which ions are retarded by negative dip in positive plasma potential. The dip value is such as to make equal losses of ions and of electrons out of the plasma (electron temperature $T_{eh}$~50 keV). Outside the ECR volume, ions are accelerated by electric field toward the chamber walls; the presheath voltage drop is free parameter in our calculations set to 2.5 V (~0.5 $T_{ec}$, where $T_{ec}$~5 eV is estimated electron temperature outside the ECR surface).

The presheath electric field is calculated in 2D approximation by *POISSON* code [10]. To do this, we approximate ECR zone by cylindrically symmetric surface and consider it as an equipotential with the fixed voltage (+2.5 V) in respect to the chamber wall potential.

Ions in the model undergo elastic ion-ion and electron-ion collisions taking into account the electron density gradients, ionizing collisions and charge-change collisions with atoms. Ions that hit the source chamber walls are neutralized and returned back with some probability depending on their surface-sticking coefficients. Energy of the neutralized particles is dependent on the wall material, mass and initial energy of impinging ions. Those ions that leave the chamber through the extraction aperture of 1 cm diameter at the extraction side of the source are considered as extracted ions.

Calculations are done in two steps: target plasma preparation and calculations with injection of (1+) ions into the plasma. First, we calculate the source dynamics with ECR discharge in pure helium gas. The DECRIS-SC2 source magnetic structure and dimensions are as follows: magnetic field at the source axis at injection and extraction are $B_{extr}$=1.36 T and $B_{inj}$=1.97 T respectively, minimal field $B_{min}$=0.49 T, hexapole field $B_{hex}$=1.1 T at the source walls, source length is 28 cm, and the source chamber diameter is 7.4 cm. The ECR zone length is 7.6 cm for the resonance magnetic field of 0.67 T calculated for 50 keV electrons.

The helium plasma parameters are obtained for the fixed magnetic fields, potential dip value of 0.1 V and gas flow of 0.6 particle-mA. At that, extracted currents of helium ions are 0.2 mA for $He^{1+}$ and 0.18 mA for $He^{2+}$, less than typical currents for helium discharges in ECRIS. The reason for small extracted currents is that such potential dip value is a rather small for helium plasmas, with the resulting poor confinement of helium ions. Mean electron density for helium plasma is calculated as $4\times10^{11}$ cm$^{-3}$, and mean charge state of helium ions inside the dense plasma is 1.6.

Only one set of helium plasma parameters is used in our simulations, with no attempts to vary the gas flow and potential dip values. The results of simulations are stored as spatially-resolved arrays of electron density, values of $n_i Z^2 = \sum_Q n_{iQ} Q^2$ ($n_{iQ}$ is ion density for helium ions in the charge state Q), of the ion temperature and of the neutral helium density.

Ion temperatures for singly and doubly charged helium ions are around 0.06 eV, being defined by the selected potential dip of 0.1 V. Neutral helium energy is around 0.03 eV, larger than the room temperature due to atom heating after collisions of helium ions with the walls.

Neutral helium density inside the ECR zone is calculated to be equal to $7.5\times10^{10}$ cm$^{-3}$, only slightly less than the gas density outside the zone of $8\times10^{10}$ cm$^{-3}$; gas burn-out inside the dense plasma is not pronounced for relatively fast helium atoms with low ionization rates.

Spatial distribution of He$^{1+}$ ions inside the plasma chamber is shown in Fig.1 as a slice in y-z plane. There, ion density is shown in false colors, with blue color corresponding to the higher values.

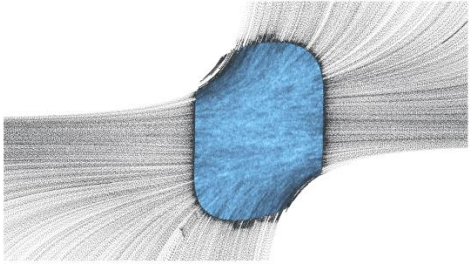

Fig.1 Spatial distribution of He$^{1+}$ ion density in y-z plane.

The distribution inside the ECR zone is uniform, with fast drop outside the zone due to ion acceleration in presheath. Distribution for He$^{2+}$ ions is basically the same as for He$^{1+}$. The prepared arrays of plasma parameters are used by the standalone charge-breeding code.

## 3. Breeding processes

Simulations of the (1+)→(n+) process are done in approximation of negligible densities of the injected ions inside the source, with no changes in the parameters of the target helium plasma. This is a rather strong simplification; in experiments, it is observed that the target ion currents are changed significantly when (1+) beam is injected into breeder [3]. It is shown in [6] that heavy highly charged ions are accumulating inside the potential trap of ECRIS plasma and change the electron life time there noticeably, with corresponding decrease in the potential dip value. At the moment, we neglect such influence, and consider the limiting case of very small current of the injected (1+) ions; implicitly, we reflect the injected beam impact on the target plasma by selecting the small value of the potential dip. In this approximation, we consider collisions of the injected ions only with ions and atoms of helium gas, neglecting inter-scattering of the injected particles.

Ion injection into the source is done along the axis from injection side. We simulate $^{23}$Na$^{1+}$, $^{39}$K$^{1+}$ and $^{85}$Rb$^{1+}$ ions. Radial size of the injected beam is small (0.1 mm), initial ion velocities are directed along the local magnetic field, and the initial energy of ions is a free parameter.

a

b

c

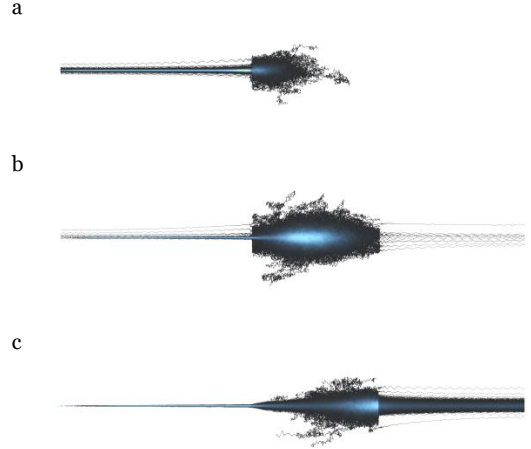

Fig.2. Trajectories of Na$^{1+}$ ions with initial energy of a) 2.6 eV, b) 5 eV and c) 7 eV.

After injection, singly charged ions are decelerated by presheath electric field while traveling toward the dense parts of helium plasma. If the initial energy $E_o$ is large enough, ions enter the plasma with energy ($E_o$-2.5) V and are elastically scattered in collision with helium ions and electrons there. Ion is captured in the plasma if after losing certain amount of energy in collisions with helium ions, its energy becomes be less than the potential dip value.

In Fig.2, trajectories of Na$^{1+}$ ions are shown inside the source for different initial energies of ions: small value of 2.6 eV represents the case when ions are stopped close to the injection side of ECR plasma, for intermediate case of 5 eV ions are captured mainly in center of the plasma, and for 7 eV most of the singly charged ions pass through the plasma and are lost at the extraction side of the source.

Due to fast rate of ion-ion collisions, the captured ions are in thermal equilibrium with helium ions inside the plasma. After ionization into the higher charge states, losses of ions out of the potential barrier trap due to diffusion in velocity space caused by ion-ion and electron-ion collisions are relatively small. Ion loss rates are almost exclusively determined, according to our calculations, by charge-change collisions with helium atoms.

Rates of charge-change collisions of ions with helium atoms are calculated following the Langevin semiclassical scaling [11], $k_{ce} = 2\pi e Q \sqrt{\frac{\alpha_{He}}{\mu}}$, where $\alpha_{He}$ is the helium static dipole polarizability (0.208 Å$^3$), μ is reduced mass $m_i m_{He}/(m_i+m_{He})$, and Q is the ion charge state. For $^{23}$Na singly charged ions, the rate is $k_{ce}=6\times10^{-10}$ cm$^3$/sec.

Only those ions experience the charge-change collisions with difference in ionization potentials I(Q)-I(Q-1)>24.6 eV, where

24.6 eV is the ionization potential of helium atom. After charge-change collisions, total kinetic energy release (KER) is calculated as either equal to the potential energy difference or to 10 eV if this difference is larger than this limiting value. Only single-electron transfer processes are taken into account. The KER is distributed between the colliding partners as prescribed by the momentum conservation; in most cases this release results in such energization of the charge-changed ion that it leaves the potential trap of plasma. Those ions are either extracted after hitting the extraction aperture or lost at the walls. We note here that the heavier is ion, the less energy it gains after experiencing the charge-exchange collision with helium atom due to kinematics of the process.

By default we consider that ions after hitting the walls are not recycled; there is an option in the code that neutralized particles are returned back into the source chamber with some probability, depending on the place of interaction with the walls.

Ionization rates for injected ions inside the ECRIS plasma are taken from datasets in [12] for the electron temperature of 50 keV. The double ionization rates are calculated according to the scaling suggested in [13]. Ion recombination processes are neglected.

Efficiency of ion breeding depends on initial energy of injected ions as it is shown in Fig.3. Here, extracted fractions for $Na^{7+}$ (left scale, black open squares) and $Na^{1+}$ (right scale, blue circles) ions are shown as function of the injected ion energy.

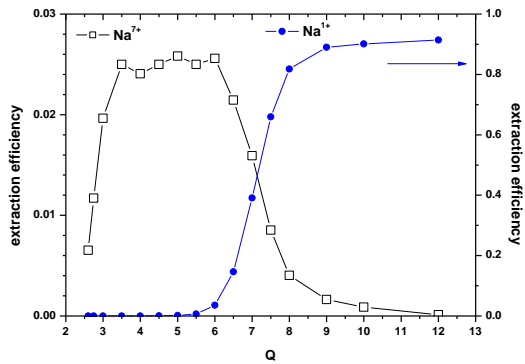

Fig.3 Efficiency of $Na^{7+}$ (left scale) and $Na^{1+}$ (right scale) ion extraction as function of the ion injection energy.

Ions are not extracted if their initial energy is less than the presheath voltage drop (2.5 V); all injected singly charged ions are reflected and lost at the injection side of the source. When ions are energetic enough to reach the ECR-limited dense plasma region, breeding efficiency first increase in course of deeper penetration of singly charged ions into the plasma followed by decrease in the fraction of ions lost at the injection side, then saturates at some level (~2.5% for $Na^{7+}$ ions), and then steadily decreases while more and more injected singly charged ions pass through the plasma without being captured there (Fig.2c).

Absolute value of injection energy threshold for ion breeding depends on sum of sheath and presheath voltages and cannot be directly compared to the experimental data presented in [2, 3, 15]. Qualitatively, the behavior of breeding efficiencies is reproduced in our calculations. For other studied elements, K and Rb, the dependencies are essentially the same as in Fig.3.

Efficiencies for extracting the ions in charge state Q for Na, K and Rb are shown in Fig.3. Initial energies of ions are selected equal to 5 eV, close to the optimum.

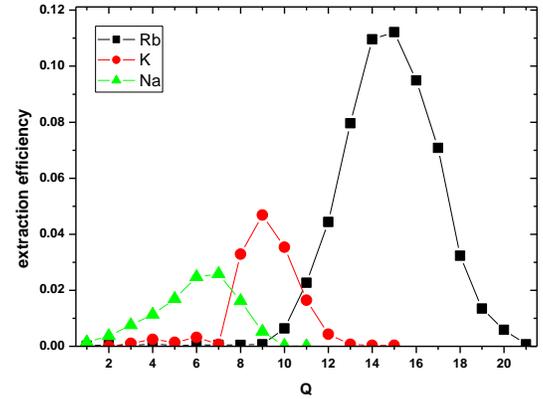

Fig.4 Extraction efficiencies for Na, K and Rb ions as function of charge state Q.

The distributions are peaked at 7+, 9+ and 15+ charge states for Na, K and Rb respectively. Results are consistent with experimental data [2,3,14,15] both in the absolute efficiencies and in shape. The heavier is the element, the larger is its maximal breeding efficiency. Total extraction efficiencies are obtained by summing fluxes into extraction aperture of ions in all charge states. The efficiencies increase with increasing the mass of injected element: for Rb this value is 60% of all injected ions, decreasing to 15% for K and 11% for Na. Diffusion of heavy ions toward the peripheral parts of the plasma is relatively slow and most of the ions are lost along the source axis into the extraction aperture.

Highly charged ions are extracted almost exclusively after charge-exchange collisions: for Na ions with initial energy of 5 eV, 90% of the extracted (7+) ions are charge-exchanged. Most of the lowly charged ions are lost without charge changing, 88% of the $Na^{2+}$ extracted ions did not experience it but overcame the potential barrier due to ion-ion and electron-ion collisions.

Time to reach equilibrium in the extracted ion currents is a rather long for highest charge states. These values are of special importance for breeding the radioactive elements and contribute into the total efficiency of breeders for short-lived isotopes. Time evolution of the extracted Na ion currents after starting the beam injection is shown in Fig.5. Pulses of $Na^{7+}$, $Na^{4+}$ and $Na^{2+}$ ions are shown there; currents of lowly charged ions are saturated in less than 5 ms, $Na^{7+}$ current equilibrates at 50 ms. Different processes are responsible for saturation for lowly and highly charged ions: ionization into the higher charge state is the most important channel of ion losses for e.g. $Na^{2+}$ and $Na^{4+}$ ions, while losses of $Na^{7+}$ ions are mainly defined by charge-exchange process. To see the influence of the charge-exchange, we plot time structure of $Na^{7+}$ pulse after starting the ion injection with setting the charge-exchange rate to zero, $k_{ce}=0$, shown in Fig.5 with the grey curve. No saturation in current is seen without charge-exchange in the 100 ms of the investigated time interval, and extracted current is substantially higher than in the default settings. For the longer times, the current of $Na^{7+}$ starts to saturate and even decreases in comparison with the "charge-exchange-on" settings, mainly because of efficient ionization into the higher charge states. With no accounting for the charge-exchange process, maximum in the extracted Na ion charge-state distribution shifts to 9+ and equals to ~15% for $Na^{9+}$ and $Na^{10+}$, order of magnitude larger compared to the data in Fig.3 and 4.

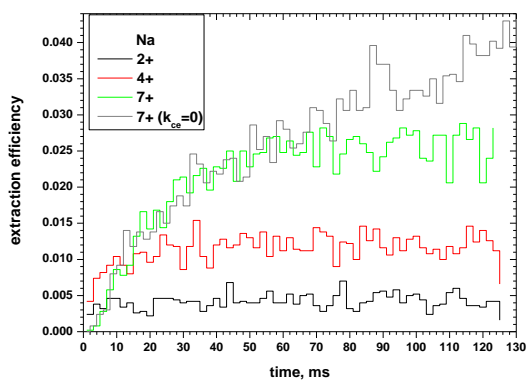

Fig.5 Time evolution of the extracted Na ions after starting the ion injection. Injection energy is 5 eV.

Only 10% of all injected Na ions are extracted in the conditions of no recycling of particles on the walls. The total efficiency is increased if allowing the back-flows of the neutralized sodium atoms into the plasma volume. The situation with setting the recycling coefficient to 100% everywhere except the extraction aperture is presented in Fig.6, where charge-state distributions of the extracted Na ions are shown with (red) and without (green) the ion recycling. Maximum in the extracted ion efficiency is increased to 6% for $Na^{7+}$ with no strong changes in the distribution shape except the increased fluxes of singly charged Na ions.

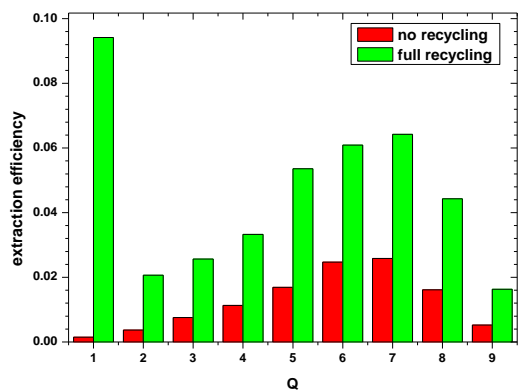

Fig.6 Charge state distributions of the extracted Na ions with and without the ion recycling on the walls.

From practical point of view, this is an indication of importance to control the back-flows of recycled particles in charge breeders by using e.g. hot screens in the chamber. Estimations of the wall recycling conditions can be done by observing the currents of singly charged ions that are high even at the relatively low injection energies.

NAM-ECRIS model is applied for simulations of charge breeding with using the helium plasma in DECRIS-SC2 source as a target. Our basic approximation is that the injected ion current is so small that its influence on the helium plasma parameters is negligible. Qualitative agreement with experimental data is obtained for breeding efficiencies and charge state distributions of the extracted ions. Charge-exchange process defines the breeding efficiency for highest charge states of ions; times to reach equilibrium between the ion injection and losses are mainly determined by the charge-exchange losses. Ion recycling on the walls strongly influences the breeding efficiency.